\documentclass{sig-alternate}
\usepackage{color}
\begin{document}

\newtheorem{definition}{Definition}

\newcommand{\CDIRECT}{d_A}
\newcommand{\CFUT}{t_A}
\newcommand{\REV}{r_A}
\newcommand{\LENGTH}{t}
\newcommand{\CREMV}{v_A(i)}
\newcommand{\CSETUP}{u_A}
\newcommand{\DIFF}{\theta_A(i,t)}
\newcommand{\DIFFA}{\theta_A(a,t)}
\newcommand{\CEXEC}{e_A(t)}
\newcommand{\CEFF}{f_F(i)}
\newcommand{\CREMVF}{v_F(i)}
\newcommand{\CREMVX}{v_x(i)}
\newcommand{\ECONOMICS}{q_A}

%%% Mittelwert-Notation mit Quer
\newcommand{\MREMV}{\bar{v}_A(\TYPE)}
\newcommand{\MREMVX}{\bar{v}_x(\TYPE)}
\newcommand{\MEXEC}{\bar{e}_A(t)}
\newcommand{\MEXECX}{\bar{e}_x(t_x)}
\newcommand{\MEFF}{\bar{f}_F(\TYPE)}
\newcommand{\MSETUP}{\bar{u}_A}
\newcommand{\MSETUPX}{\bar{u}_x}
\newcommand{\MREMVF}{\bar{v}_F(\TYPE)}

%%% Defect classes and types
\newcommand{\TYPE}{\tau_i}
\newcommand{\TYPEJ}{\tau_j}
\newcommand{\PTYPE}{p_i(\TYPE)}
\newcommand{\CLASSES}{C}
\newcommand{\CLASS}{c}
\newcommand{\ECLASS}[1]{E_{p_i^c}(#1)}
\newcommand{\DDT}{de\-fect-de\-tec\-tion tech\-nique}
\newcommand{\DIFFT}{\theta_A(\TYPE,t)}

\newcommand{\PCLASS}{q_i(\CLASS)}

%%% E-Notation derzeit nicht genutzt
\newcommand{\EREMV}{E_\alpha(i,\CREMV)}
\newcommand{\EEFF}{E_\sigma(i,\CEFF)}
\newcommand{\EREMVF}{E_\alpha(\CLASS,\CREMVF)}
\newcommand{\EREMVX}{E_\alpha(i,\CREMVX)}

%%% o2-Fallstudie
\def\O2{$\textnormal{O}_2$}

%%% MOST-Fallstudie
\def\NM{Net\-work\-Master}
\def\AF{{\sc Auto\-Focus}}
\def\REG{Reg\-is\-try\-Mgr}
\def\REQ{Re\-quest\-Mgr}
\def\SCC{system configuration check}

%
% --- Author Metadata here ---
\conferenceinfo{ISSTA'06,}{July 17--20, 2006, Portland, Maine, USA.}
\CopyrightYear{2006} % Allows default copyright year (2000) to be over-ridden - IF NEED BE.
\crdata{1-59593-263-1/06/0007}  % Allows default copyright data (0-89791-88-6/97/05) to be over-ridden - IF NEED BE.
% \permission{\copyright ACM, 2006. This is the author's version
% of the work. It is posted here by permission of ACM for your personal
% use. Not for redistribution. The definitive version was published in
% the proceedings.}
% --- End of Author Metadata ---

\title{A Model and Sensitivity Analysis of the Quality Economics of 
       Defect-Detection Techniques}
%
% You need the command \numberofauthors to handle the "boxing"
% and alignment of the authors under the title, and to add
% a section for authors number 4 through n.
%
% Up to the first three authors are aligned under the title;
% use the \alignauthor commands below to handle those names
% and affiliations. Add names, affiliations, addresses for
% additional authors as the argument to \additionalauthors;
% these will be set for you without further effort on your
% part as the last section in the body of your article BEFORE
% References or any Appendices.

\numberofauthors{1}
%
% You can go ahead and credit authors number 4+ here;
% their names will appear in a section called
% "Additional Authors" just before the Appendices
% (if there are any) or Bibliography (if there
% aren't)

% Put no more than the first THREE authors in the \author command
\author{
%% e-mail address with \email.
\alignauthor Stefan Wagner\\
       \affaddr{Institut f\"ur Informatik}\\
       \affaddr{Technische Universit\"at M\"unchen}\\
       \affaddr{Boltzmannstr.\ 3, D-85748 Garching b.\ M\"unchen, Germany}\\
       \email{wagnerst@in.tum.de}
}

\maketitle

\begin{abstract}
One of the main cost factors in software development is the detection
and removal of defects. However, the relationships and influencing
factors of the costs and revenues of defect-detection techniques
are still not well understood. This paper proposes an analytical,
stochastic model of the economics of defect detection and removal
to improve this understanding.
The model is able to incorporate dynamic as well as static techniques
in contrast to most other models of that kind. We especially analyse
the model with state-of-the-art sensitivity analysis methods
to (1) identify the
most relevant factors for model simplification and (2) prioritise
the factors to guide further research and measurements.
\end{abstract}

%A category including the fourth, optional field follows...



\section{Introduction}

The quality of a software system can be described using different
attributes such as reliability, maintainability etc. There are also various
approaches to improve the quality of software with differing emphasis
on these attributes. Constructive methods comprise one group that tries to
improve the overall development process in order to prevent the introduction
of faults. However, the prevalent approach is still to use analytical methods,
also called \DDT s, to find and remove faults. The main representatives of
this approach are tests and reviews.

An often cited estimate \cite{myers79} relates
50\% of the overall development costs to testing. Jones
\cite{jones91} still assigns 30--40\% to quality assurance and defect
removal.
Hence, \DDT s are a
promising field for cost optimisations. 
However, to be able to optimise the usage of \DDT s, we need a suitable
 economical
model first. There are some approaches that model software quality costs
but mostly on a high level of abstraction. The effects of individual faults
and the effectiveness of different \DDT s regarding these faults are not
taken into account.
Also in \cite{ntafos01} it is discussed that ``cost is clearly a central
factor in any realistic comparison but it is hard to measure, data are not
easy to obtain, and little has been done to deal with it.''
Rai et al.\ identify in \cite{rai98} mathematical models of the economics 
of software
quality assurance as an important research area. ``A better understanding
 of the costs and benefits of SQA and improvements to existing quantitative
models should be useful to decision-makers.''

\subsection{Problem}
The underlying question is how we can optimally use \DDT s to improve
the quality of software. In particular, we investigate in this paper
how the economical relationships of \DDT s and quality can be modelled
and the importance of the factors in terms of the influence on the
output and especially its variance.

\subsection{Contribution}
We propose an analytical model of the economics of \DDT s incorporating
different types of defect costs, the difficulty of finding a fault of
different techniques and the probability of failure for a fault. This
allows an evaluation of different techniques and gives a better understanding
of the relationships. The used input factors are prioritised to simplify
the model and to identify the factors that are most beneficial to be
further investigated. Furthermore, a model based on defect types is derived
to allow a simpler application on real world projects. This model could
be used to predict optimal usage of \DDT s in the future based on old
project data.

\subsection{Outline}
We start by describing software quality costs in general and our understanding
of the various cost factors in Sec.~\ref{sec:costs}. Sec.~\ref{sec:ideal}
proposes an analytical model of the economics of \DDT s that contains the costs
associated with each fault. This model is subject to a sensitivity
analysis based on the data from an older study in Sec.~\ref{sec:sensitivity}.
For the practical application of the model a simplified version 
based on defect classes
is derived in Sec.~\ref{sec:practical}.
Sec.~\ref{sec:related}
gives related work and final conclusions can be found in 
Sec.~\ref{sec:conclusions}.

\section{Software quality costs}
\label{sec:costs}

\emph{Quality costs} are the costs associated with preventing, finding,
and correcting defective work.
Based on experience from the manufacturing area \cite{juran98,feigenbaum05}
similar quality cost models have been developed explicitly for software
\cite{knox93,slaughter98, Krasner1998}.
These costs are divided into \emph{conformance} and \emph{nonconformance}
costs, also called \emph{control costs} and \emph{failure of control costs}. 
The former comprises all costs that need to be spent to build the software in
a way that it conforms to its quality requirements. This can be further
broken down to \emph{prevention} and \emph{appraisal} costs. Prevention 
costs are for
example developer
training, tool costs, or quality audits, i.\,e.~costs for means to prevent
the injection of faults. The appraisal costs are caused by the usage of 
various types of tests and reviews.

\begin{figure}[h]
  \centering \includegraphics[width=8cm]{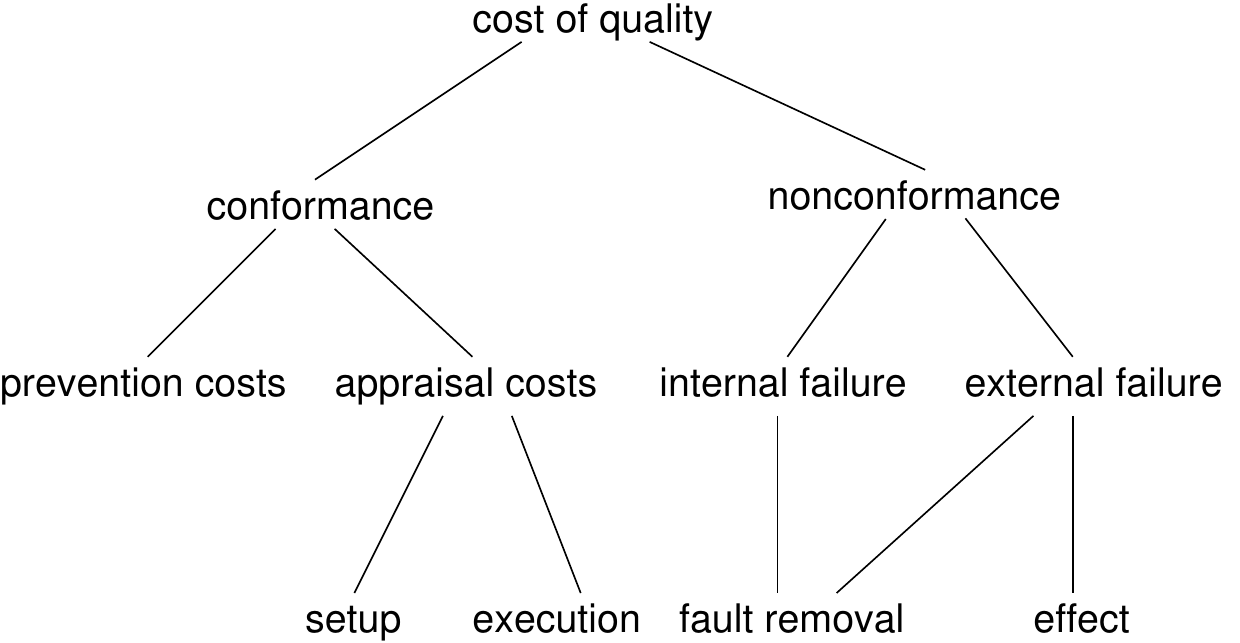}
  \caption{Overview over the costs related to quality}
  \label{fig:costs_overview}
\end{figure}

The \emph{nonconformance} costs come into play when the software does not
conform to the quality requirements. These costs are divided into
\emph{internal failure} costs and \emph{external failure} costs. The
former contains costs caused by failures that occur during development,
the latter describes costs that result from failures at the client.
A graphical overview is given in Fig.~\ref{fig:costs_overview}. Because
of the distinction between prevention, appraisal, and failure costs this
is often called \emph{PAF} model.
 
We add further detail to the PAF model
by introducing the main types of concrete costs that are important for
defect-detection techniques. Note that there are
more types that could be included, for example, maintenance costs.
However, we concentrate on a more reliability-oriented view.
The appraisal costs
are detailed to the \emph{setup} and \emph{execution} costs. The former
constituting all
initial costs for buying test tools, configuring the test environment,
and so on. The latter means all the costs that are connected to actual
test executions or review meetings, mainly personnel costs.

On the nonconformance side, we have \emph{fault removal} costs that can
be attributed to the internal failure costs as well as the external
failure costs. This is because if we found a fault and want to remove it,
it would always result in costs no matter whether caused in an internal
or external failure. Actually, there does not have to be a failure at all.
Considering code inspections, faults are found and removed that have never
caused a failure during testing. It is also a good example that the removal
costs can be quite different regarding different techniques. When a test
identifies a failure, there needs to be considerable effort spent to find
the corresponding fault. During an inspection, faults are found directly.
Fault removal costs also contain the costs for necessary re-testing and
re-inspections.

External failures also cause \emph{effect} costs. These are all further costs
with the failure apart from the removal costs. For example, \emph{compensation}
costs could be part
of the effect costs, if the failure caused some kind of damage
at the customer site. We might also include further costs such as loss of
sales because of bad
reputation in the effect costs but do not consider it explicitly
because its out of scope of this paper.

\section{Analytical Model}
\label{sec:ideal}

We describe a general, analytical model of \DDT s in the following.
It is general 
with respect to the various types of techniques it is able to analyse. 
We mainly analyse different types of testing which essentially
detect failures and static analysis techniques that reveal faults in the
code or other documents. A model that incorporates all important factors 
for these differing techniques needs to use
the universal unit of money, i.e., units such as euro or dollar.
We first describe the model and its assumptions in general, 
and then give equations
for each component of the model for a single technique and for the
combination of several techniques.

\subsection{General}

In this section, we concentrate on an ideal model of quality economics
in the sense that we do not consider the practical use of the model but
want to mirror the actual relationships as faithfully as possible. 

\subsubsection{Components}

We divide the model in three main components:
\begin{itemize}
  \item Direct costs $\CDIRECT$
  \item Future costs $\CFUT$
  \item Revenues / saved costs $\REV$
\end{itemize}
The direct costs are characterised by containing only costs that
can be directly measured during the execution of the technique. The
future costs and revenues are both concerned with the (potential) costs in
the field but can be distinguished because the future costs contain the
costs that are really incurred whereas the revenues are comprised of
saved costs.

\subsubsection{Assumptions}

The main assumptions in the model are:
\begin{itemize}
  \item Found faults are perfectly removed.
  \item The amount or duration of a technique can be freely varied.
\end{itemize}

The first assumption is often used in software reliability modelling
to simplify the stochastic models. It states that each fault detected
is instantly removed without introducing new faults. Although this is
often not true in real defect removal, it is largely independent of the
used \DDT\ and the newly introduced faults can be handled like initial
faults which introduces only a small blurring.

The second assumption is needed because we have a notion of time effort
in the model to express for how long and with how many people a technique
is used. This notion of time can be freely varied although for real
\DDT s this might not always make sense, especially when considering
inspections or static analysis tools where a certain basic effort
or none at all has to be spent. Still, even for those techniques, the
effort can be varied by changing the speed of reading, for example.

\subsubsection{Difficulty}

We adapt the general notion of the difficulty of a technique $A$ to find
a specific fault $i$ from \cite{littlewood:tse00} denoted by $\theta_A (i)$
as a basic quantity for our model. In essence, it is the probability
that $A$ does not detect $i$. Furthermore, we denote the length of
a technique application with $t_A$. With length we do not mean calendar
time but effort measured in staff-days, for example.

In the following equations
we are often interested in the case when a fault is detected at least
once by a technique. From the above we can conclude that the probability
that $A$ detects $i$ is $1-\theta_A (i)$. However, as stated above, we
have a concept of timing and effort for a technique that has to be
incorporated in the difficulty. Hence, with $t_A$ denoting the
effort spent for $A$,
the probability that $i$ is at least detected once is $1 - \theta_A (i,t_A)$.

\subsubsection{Defect Propagation}

\begin{figure}[h]
\begin{center}
  \includegraphics[width=.4\textwidth]{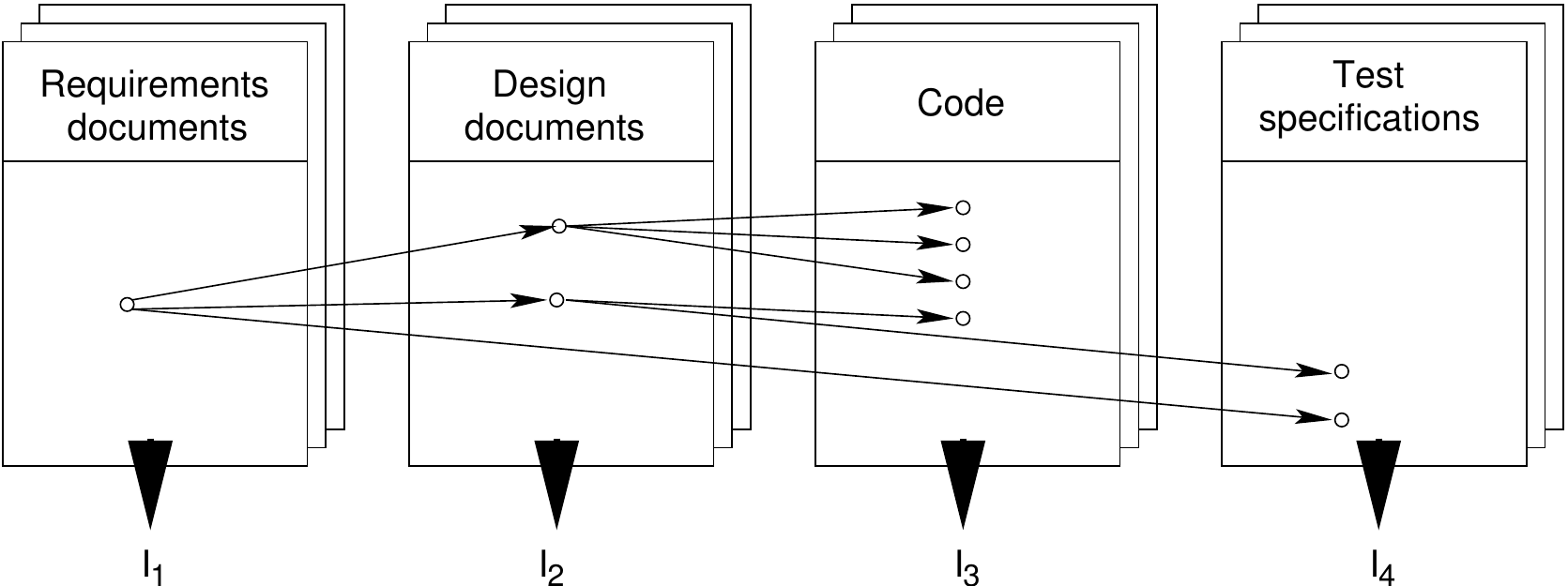}
  \caption{How defects propagate over documents}
  \label{fig:defect_propagation}
\end{center}
\end{figure}

A further aspect to consider is that the defects occurring during
development are not independent. There are various dependencies
that could be considered but most importantly there is dependency
in terms of propagation.
Defects from earlier phases propagate to later phases and over
process steps. We actual do not consider the phases to be the important
factor here but the document types. In every development process there
are different types of documents, or artifacts, that are created.
Usually, those are requirements documents, design documents, code,
and test specifications. Then one defect in one of these documents
can lead to none, one, or more defects in later derived documents.
A schematic overview is given in Fig.~\ref{fig:defect_propagation}.

We see that a requirements defect can lead to several defects
in design documents as well as test specifications. The design
defects can again propagate to the code and to (glass-box)
test specifications. For each document type $k$ we have the
set of defects $I_k$ and hence the total set of defects $I$
is $I = \bigcup{I_k}$. Furthermore, for each defect, we also
look at its predecessor defects $R_i$. For the model this 
has the effect that a defect can only be found by a technique
if neither the defect itself nor one of its predecessors
was detected by an earlier used technique.

%%%%%%%%%%%%%%%%%%%%%%%%%%%%%%%%%%%%%%%%%%%%%%%%%%%%%%%%%%%%%%%
\subsection{Equations}
\label{sec:equations_ideal}

We give an equation for each of the three components with respect
to single \DDT s first and later for a combination of techniques.

\subsubsection{Direct Costs}

\begin{figure}[h]
  \begin{center}
    \includegraphics{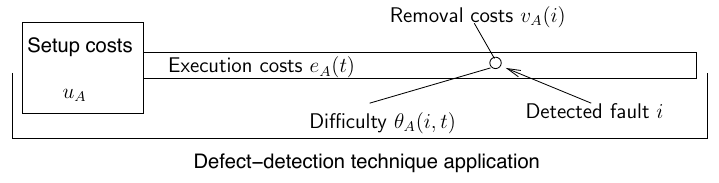}
  \end{center}
  \caption{The components of the direct costs
  \label{fig:components}}
\end{figure}

The direct costs are those costs that can be directly measured from
the application of a \DDT . They are dependent on the length $\LENGTH$
of the application.
Fig.~\ref{fig:components} shows systematically
the components of the direct costs.

From this we can derive the following equation containing the
three cost types for a technique.

\begin{equation}
  \label{eq:direct}
  \CDIRECT = \CSETUP + \CEXEC
             + \sum_{i}{
             (1 - \DIFF) \CREMV},
\end{equation}
where $\CSETUP$ are the setup costs, $\CEXEC$ the execution costs, and
$\CREMV$ the fault removal costs specific to that technique. Hence, we
have for a technique its fixed setup costs, execution costs depending
on the length of the technique and for each fault in the software
removal costs if the technique is able to find it.

\subsubsection{Future Costs}

In case we were not able to find defects, these will result in costs
in the future. We divide these costs into the two parts fault removal
costs in the field $\CREMVF$ and failure effect costs $\CEFF$. The
latter contain all support and compensation costs as well as annoyed
customers as far as possible.

\begin{equation}
\label{eq:future}
  \CFUT = 
  \sum_i{\pi_i \DIFF (\CREMVF + \CEFF)},
\end{equation}
where $\pi_i = P$(fault $i$ is activated by
randomly selected
input and is detected and
fixed) \cite{littlewood:tse00}. Hence, it describes the probability that 
the defect leads to a failure
in the field.

\subsubsection{Revenues}

We do not only have costs with \DDT s but also revenues. These revenues
are essentially saved future costs. With every fault that we find in-house
we avoid higher costs in the future. Therefore, we have the same cost
categories but look at the faults that we find instead of the ones we
are not able to detect.

\begin{equation}
  \label{eq:saved}
  \REV =
  \sum_i{\pi_i (1 - \DIFF)(\CREMVF + \CEFF)}
\end{equation}

\subsubsection{Combination}

Typically, more than one technique is used to find defects.
The intuition behind that is that they find (partly) different defects.
These dependencies 
are often
ignored when the efficiency of \DDT s is analysed. Nevertheless, this
has a huge influence on the economics and efficiency. In our view, the
notion of diversity of techniques from Littlewood et al.~\cite{littlewood:tse00}
is very useful in this context. The covariance of the difficulty functions
of faults describes the similarity of the effectiveness regarding fault
finding. 
We already use the difficulty functions in the
present model and therefore are able to express the diversity implicitly.

For the direct costs it means that we sum over all different
applications of \DDT s. We define that
$X$ is the ordered set of the applied \DDT s. In each application we
use Eq.~\ref{eq:direct} with the extension that we not only take the
probability that the technique finds the fault into account but 
also that the ones
before have not detected it. Here also the defect propagation
needs to be considered, i.e., that not only the defect itself has
not been detected but also its predecessors $R_i$.

\begin{equation}
\begin{split}
\label{eq:direct_total}
  d_X = \sum_{x \in X}{\biggl[ u_x + e_x(t_x)} + 
        \sum_i{\Bigl( (1 - 
         \theta_x(i,t_x))}\\
  \phantom{d_X = } \prod_{y < x}{\theta_y(i,t_y)}
                         \prod_{j \in R_i}{\theta_y(j,t_y)}\Bigr) 
         \CREMVX \biggr]
\end{split}
\end{equation}

The total future costs are simply the costs of each fault with the
probability that it occurs and all techniques failed in detecting it
and its predecessors.

\begin{equation}
\begin{split}
\label{eq:future_total}
  t_X = \sum_i{\biggl[ \pi_i \prod_{x \in X}{\theta_x(i,t_x)}}\\
\phantom{t_X} \prod_{y < x}{\prod_{j \in R_i}{\theta_y(j,t_y)}}
        (\CREMVF + \CEFF) \biggr] 
\end{split}
\end{equation}

The equation for the revenues uses again a sum over all technique
applications. In this case we look at the faults that occur, that are
detected by a technique and neither itself nor its predecessors
have been detected by the earlier
applied techniques.

\begin{equation}
\begin{split}
\label{eq:revenues_total}
  r_X = \sum_{x \in X}{\sum_i{\biggl[ \Bigl( \pi_i (1 - \theta_x(i,t_x)) 
         \prod_{y < x}{\theta_y(i,t_y)}}}\\
         \prod_{j \in R_i}{\theta_y(j,t_y)}
         \Bigr) 
         \bigl( 
         \CREMVF + \CEFF \bigr)  \biggr] 
\end{split}
\end{equation}

\subsubsection{ROI}

One interesting metric based on these values is the return on investment
(ROI) of the \DDT s. If we look at the total ROI we have to use 
Eqns.~\ref{eq:direct_total}, \ref{eq:future_total}, and 
\ref{eq:revenues_total}
for the calculation.

\begin{equation}
  \mbox{ROI} = \frac{r_X - d_X - t_X}{d_X + t_X}
\end{equation}

This metric is suitable for a single post-evaluation of the quality
assurance of a project. However, it alone cannot give an answer whether the 
effort was cost-optimal.

\subsection{Forms of the Difficulty Functions}
\label{sec:func_forms}

The notion of \emph{difficulty} of the defect detection is a very
central one in the described model. As mentioned this notion is based
on an idea from \cite{littlewood:tse00}. However, the original difficulty
functions have no concept of time or effort spent but only of one
usage or two usages and so on. To be able to analyse and optimise the
spent effort for each technique, we need to introduce that additional
dimension in the difficulty functions, i.e., the functional form depending
on the spent effort. Actually, the equations given for the model above
already contain that dimension but it is not further elaborated.
This gap is closed in the following.

Firstly, we do not have sufficient data to give an empirically founded
basis for the forms of the difficulty functions. Nevertheless,
we can formulate hypotheses to identify the most probable distributions
for different defects. Secondly, keep in mind that a difficulty function
is defined for a specific \DDT\ detecting a specific defect. That means
that each defect can have distinct distribution for each possible technique.

\subsubsection{Exponential Function}

The function that most obviously models the process under investigation
is an exponential function. 
The intuition is that with more effort spent the difficulty decreases,
i.e., the probability of detecting that defect increases. However, with
increasing effort the rate of difficulty reduction slows down. The
defect detection gets more and more complicated when the ``obvious''
cases all have been tried.

For this we use a function similar to the density function of an exponential
distribution:
\begin{equation}
\theta(i,t) = \left\{ \begin{array}{ll}
                        \lambda e^{-\lambda t} & \mbox{if $t>0$} \\
                        1 & \mbox{otherwise}
                      \end{array}
              \right. ,
\end{equation}
with $\lambda$ being a parameter that is determined from empirical data
from the technique and the defect. It is the inverse of the mean value
of the empirically measured difficulty.

\subsubsection{Linear and Constant Function}

The linear difficulty function
models the intuition that there is a steady decrease in difficulty.
A review might be an example that employs such a behaviour. The more
I read, the higher the possibility that I detect that specific defect.
The function can be formulated as follows:
\begin{equation}
\label{eq:linear}
\theta_A(\tau_i, t) = m t + 1,
\end{equation}
where $m$ is the (negative) slope of the straight line.

The constant function constitutes a special case of the
linear form of the difficulty functions. In this case the spent effort
does not matter because the difficulty of detecting
the defect is always the same. 
The intuitive explanation for this functional form is best explained
using the example of a static analysis tool. These tools often use
bug patterns specific for a language and thereby identify code sections
that are critical. When searching for a specific bug pattern it is
of no importance how much effort is spent but if the tool is not able
to detect a specific pattern -- or only in seldom cases -- the
probability of detection does not change. We can also use this
distribution to model that a specific technique $A$ cannot detect a
specific defect $i$ by specifying that $\theta_A(i,t) = 1$ for all $t$.

\subsubsection{Sigmoid Function}

For our purposes it is sufficient to see the
sigmoid function as a variation of the
exponential function. Its graph has an
\emph{S}-like shape and hence one local minimum and one local maximum.
In this special case we actually
use a complementary sigmoid function to get a turned \emph{S}.

In contrast to the exponential function, the sigmoid function
models the intuition that in the beginning it is hard to detect a
specific defect and the difficulty does decrease only slowly. However,
when a certain amount of effort is spent, the rate increases and chance
of detecting the defect increases significantly until we reach a point
of satisfaction -- similar to the exponential function -- where
additional effort does not have a large impact. This distribution is
also backed by the so-called S-curve of software testing \cite{kan02}.
That S-curve aims in a slightly different direction but also shows that
early in testing only a limited number of failures are revealed, then
the detection rate increases until a plateau of satisfaction is reached.

\subsection{Discussion}
\label{sec:discussion}

The model so far is not suited for a practical application in a
company as the quantities used are not easy to measure. Probably, we
are unable to get values for $\theta$ of each fault and \DDT .
Also the somehow fixed and distinct order of techniques is not
completely realistic as some techniques may be used in parallel or
only some parts of the software are analysed.
However, in a more theoretical setting we can already use the model 
for important tasks including
sensitivity analysis to identify important input factors.

Another application can be to analyse which techniques influence which
parts of the model. For instance, in the automatic derivation of
test cases from explicit behaviour models (model-based testing) is
a relatively new technique for defect detection. This technique can
be analysed and compared with traditional, hand-crafted test techniques
based on our model. Two of the factors are obviously affected by
model-based testing: (1) the setup costs are considerably higher than
in hand-crafted tests because not only the normal test environment
has to be set up but also a formal (and preferably executable)
behaviour has to be developed. On the contrary, the execution cost
per test case is then substantially smaller because the generation
can be automated to some extent and the model can be used as an oracle
to identify failures. Further influences on factors like the difficulty
functions are not that obvious but need to be analysed. This example
shows that the model can help to structure the comparison and analysis
of defect-detection techniques.

\section{Sensitivity Analysis}
\label{sec:sensitivity}

Every newly proposed mathematical model should be subject to various analyses.
Apart from the appropriateness of the model to the modelled reality and
the validity of estimates and predictions, the dependence of the output
on the input parameters is of interest. The quantification of this
dependence is called \emph{sensitivity analysis}. Local sensitivity
analysis usually computes the derivative of the model response with
respect to the model input parameters. More generally applicable is
global sensitivity analysis that apportions the variation in the
output variables to the variation of the input parameters. We base
the following description of the global sensitivity analysis we use
mainly on \cite{saltelli04}.

\subsection{Settings and Methods}

Sensitivity analysis is
the study of how the uncertainty in the output of a model
can be apportioned to different sources of uncertainty in the
model input. Still, what do we gain by that knowledge? There are
various questions that can be answered by sensitivity analysis. As
pointed out in \cite{Saltelli2004} it is important to specify its
purpose beforehand. In our context two settings are of most interest:
(1) factors priorisation and (2) factors fixing.

\subsubsection{Factors Priorisation} 
The most important factor is the one that
would lead to the greatest reduction in the variance of the output if
fixed to its true value. Likewise the second most important factor can
be defined and so on.
The ideal use for the Setting FP is for the prioritisation of research and
this is one of the most common uses of sensitivity analysis in general.
Under the hypothesis that all
uncertain factors are susceptible to determination, at the same cost per
factor, Setting FP allows the identification of the factor that is most
deserving of better experimental measurement in order to reduce the target
output uncertainty the most. In our context, that means that we can
determine the factors that are most rewarding to measure most precisely.

\subsubsection{Factors Fixing}
This setting is similar to factors priorisation but still has a slightly
different flavour. Now, we do not aim to prioritise research in the
factors but we want to simplify the model. For this we try to find the
factors that can be fixed without reducing the output
variance significantly. For our purposes this means that we can fix the input factor
at any value in its uncertainty range without changing the outcome
significantly.

\subsubsection{FAST}
There are various available methods for global sensitivity analysis.
The \emph{Fourier amplitude sensitivity test (FAST)} is a commonly
used approach that is based on Fourier developments of the output
functions. It also allows an ANOVA-like decomposition of the model
output variance. In contrast to correlation or regression coefficients,
it is not dependent on the goodness of fit of the regression model.
The results give a quantification of the influences of the parameters,
not only a qualitative ranking as the Morris method, for example.

With the latest developments of this method, it is able not only
to compute the first-order effects of each input parameter but also
the higher-order and total effects. The first order effect is the
influence of a single input parameter on the output variance, whereas
the total effects also capture the interaction between input parameters.
This is also important for the different settings as the first-order
effects are used for the factors priorisation setting, the total-order
effects for the factors fixing setting.

\subsubsection{SimLab}
We use the sensitivity analysis tool \emph{SimLab} \cite{Simlab} for 
the analysis. Inside the tool we need to define all needed input
parameters and their distributions of their uncertainty ranges. For
this, different stochastic distributions are available.
The tool then generates the 
samples needed for the analysis. This sample data is can be read from
a file into the model -- in our case a Java program -- that is
expected to write its output into a file with a specified format.
This file is read again from SimLab and the first-order and total-order
indexes are computed from the output.

\subsection{Input Factors and Data} 

We describe the analysed scenario, factors and data needed
for the sensitivity analysis in the following. The distributions
are derived from the survey \cite{wagner:tumi06,Wagner2006}. We base
the analysis on an example software with 1000 LOC and with 10--15
faults. The reason for the small number of faults is the increase
in complexity of the analysis for higher numbers of faults.

\subsubsection{Techniques}
We have to base the sensitivity analysis on common or \emph{average}
distributions of the input factors. This also implies that we use a
representative set of \DDT s in the analysis. 
We choose seven commonly used \DDT s and
encode them with numbers: requirements inspection (0), design
inspection (1), static analysis (2), code inspection (3), (structural)
unit test (4), integration test (5), and (functional) system test (6).
As indicated we assume that unit testing is a structural (glass-box)
technique, system testing is functional (black-box), and integration
testing is both. The usage of those seven techniques, however, does
not imply that all of them are used in each sample as we allow the
effort $t$ to be null.

\subsubsection{Additional Factors}
To express the defect propagation concept of the model we added the
additional factor
$\rho$ as the number of predecessors. The factor $c$ represents the
the defect class meaning the type of artifact the defect is contained in.
This is important for the decision whether a certain technique is capable
to find that specific defect at all. The factor
$\phi$ encodes the form of difficulty function that is used for a specific
fault and a specific technique. We include all the forms presented above
in Sec.~\ref{sec:func_forms}.
The sequence $s$ of techniques determines the order of execution of the
techniques. We allow several different sequences including nonsense
orders in which system testing is done first and requirements inspections
as the last technique. Finally, the average labour costs per hour
$l$ is added because it is not explicitly included in the model equations
from Sec.~\ref{sec:equations_ideal}. Note, that we excluded the effect
costs from the sensitivity analysis because we have not sufficient data
to give any probability distribution.

\subsection{Results and Observations}

This section summarises the results of applying the
FAST method for sensitivity analysis on the data from the example
above and discusses observations. The analysed output factor is
the return-on-investment (ROI).

\subsubsection{Abstract Grouping}

We first take an abstract view on the input factors and group them
without analysing the input factors for different techniques separately.
Hence, we only have 11 input factors that are ordered with respect to
their first and total order indexes in Tab.~\ref{tab:ideal_abstract}.
The first order indexes are shown on the left, the total order indexes
on the right.

\begin{table}[htbp]
\caption{The first and total order indexes of the abstract grouping}
\begin{center}
\begin{tabular}{|l|r|l|r|}
\hline
$c$ & 0.4698 & $c$ & 0.8962 \\ \hline
$t$ & 0.1204 & $v_f$ & 0.4473 \\ \hline
$\bar{\theta}$ & 0.0699 & $\bar{\theta}$ & 0.4255 \\ \hline
$v_f$ & 0.0541 & $u$ & 0.3916 \\ \hline
$\phi$ & 0.0365 & $t$ & 0.3859 \\ \hline
$u$ & 0.0297 & $\phi$ & 0.2888 \\ \hline
$v$ & 0.0264 & $\rho$ & 0.2711 \\ \hline
$\rho$ & 0.0256 & $v$ & 0.2546 \\ \hline
$\pi$ & 0,0158 & $\pi$ & 0,2068 \\ \hline
$s$ & 0,0083 & $s$ & 0,1825 \\ \hline
$l$ & 0,0010 & $l$ & 0,1489 \\ \hline
\end{tabular}
\end{center}
\label{tab:ideal_abstract}
\end{table}

The first order indexes are used for the factors priorisation setting.
We see that the types of documents or artifacts the defects are contained
in are most rewarding to be investigated in more detail. One reason
might be that we use a uniform distribution because we do not have
more information on the distribution of defects over document types.
However, this seems to be an important information. The factor that
ranks second highest is the spent effort. This approves the intuition
that the effort has strong effects on the output and hence needs to
be optimised. Also the average
difficulty of of finding a defect with a technique and the costs of
removing a defect in the field are worth to be investigated further.
Interestingly, the labour costs, the sequence of technique application,
and the failure probability in the field do not contribute strongly to
the variance in the output. Hence, these factors should not be the
focus in further research.

For the factors fixing setting, the ordering of the input factors is
quite similar. Again the failure probability in the field, the sequence
of technique application and the labour costs can be fixed without
significantly changing the output variance. The factors that definitely
cannot be fixed are again the document types, the removal costs in the
field, and the average difficulty values. The setup costs rank higher
with these indexes and hence should not be fixed.

\subsubsection{Detailed Grouping}

After the abstract grouping, we form smaller groups and differentiate
between the factors with regard to different \DDT s. The first and
total order indexes are shown in Tab.~\ref{tab:ideal_detailed} again
with the first order indexes on the left and the total order indexes
on the right.

\begin{table}[htbp]
\caption{The first and total order indexes of the detailed grouping}
\begin{center}
\begin{tabular}{|l|r|l|r|}
\hline
$c$ & 0.2740 & $c$ & 0.7750 \\ \hline
$t_1$ & 0.0601 & $\phi_4$ & 0.3634 \\ \hline
$\pi$ & 0.0528 & $t_1$ & 0.3332 \\ \hline
$\phi_4$ & 0.0492 & $\pi$ & 0.3200 \\ \hline
$\phi_1$ & 0.0391 & $v_f$ & 0.2821 \\ \hline
$v_3$ & 0.0313 & $v_3$ & 0.2802 \\ \hline
$\phi_0$ & 0.0279 & $\phi_1$ & 0.2728 \\ \hline
$\rho$ & 0.0278 & $\rho$ & 0.2706 \\ \hline
$\phi_2$ & 0.0269 & $v_1$ & 0.2574 \\ \hline
$v_f$ & 0.0252 & $s$ & 0.2524 \\ \hline
$\phi_6$ & 0.0222 & $\bar{\theta}_5$ & 0.2493 \\ \hline
$v_0$ & 0.0219 & $\bar{\theta}_0$ & 0.2312 \\ \hline
$\phi_3$ & 0.0216 & $\bar{\theta}_3$ & 0.2300 \\ \hline
$\bar{\theta}_6$ & 0.0214 & $\phi_6$ & 0.2287 \\ \hline
$v_5$ & 0.0212 & $\phi_2$ & 0.2240 \\ \hline
$\bar{\theta}_0$ & 0.0209 & $\bar{\theta}_1$ & 0.2077 \\ \hline
$s$ & 0.0208 & $v_5$ & 0.2039 \\ \hline
$\bar{\theta}_1$ & 0.0203 & $\phi_0$ & 0.1966 \\ \hline
$v_1$ & 0.0203 & $u_3$ & 0.1913 \\ \hline
$\bar{\theta}_4$ & 0.0197 & $v_0$ & 0.1907 \\ \hline
$\phi_5$ & 0.0194 & $\bar{\theta}_6$ & 0.1894 \\ \hline
$\bar{\theta}_5$ & 0.0186 & $\phi_5$ & 0.1892 \\ \hline
$t_2$ & 0.0185 & $\phi_3$ & 0.1854 \\ \hline
$\bar{\theta}_3$ & 0.0181 & $v_4$ & 0.1807 \\ \hline
$v_6$ & 0.0142 & $t_5$ & 0.1719 \\ \hline
$v_2$ & 0.0139 & $v_6$ & 0.1709 \\ \hline
$v_4$ & 0.0120 & $\bar{\theta}_4$ & 0.1707 \\ \hline
$\bar{\theta}_2$ & 0.0109 & $v_2$ & 0.1633 \\ \hline
$t_6$ & 0.0089 & $t_6$ & 0.1619 \\ \hline
$t_4$ & 0.0058 & $u_5$ & 0.1451 \\ \hline
$t_3$ & 0.0051 & $t_4$ & 0.1409 \\ \hline
$t_5$ & 0.0034 & $u_4$ & 0.1404 \\ \hline
$u_5$ & 0.0018 & $t_2$ & 0.1378 \\ \hline
$u_0$ & 0.0013 & $t_0$ & 0.1268 \\ \hline
$u_4$ & 0.0010 & $l$ & 0.1222 \\ \hline
$t_0$ & 0.0009 & $\bar{\theta}_2$ & 0.1125 \\ \hline
$u_3$ & 0.0007 & $u_6$ & 0.1122 \\ \hline
$u_6$ & 0.0007 & $u_0$ & 0.1085 \\ \hline
$u_1$ & 0.0005 & $t_3$ & 0.1053 \\ \hline
$u_2$ & 0.0005 & $u_1$ & 0.1034 \\ \hline
$l$ & 0.0002 & $u_2$ & 0.0996 \\ \hline
\end{tabular}
\end{center}
\label{tab:ideal_detailed}
\end{table}

The main observations from the abstract grouping for the factor
priorisation setting are still valid.
The type of artifact the defect is contained in still ranks highest
and has the most value in reducing the variance. However, in this
detailed view, the failure probability in the field ranks higher.
This implies that this factor should not be neglected. We also see
that for some techniques the form of the difficulty function has
a strong influence and that the setup costs of most techniques rank
low.

Similar observations can be made for the factors fixing setting and
the total order indexes. The main observations are similar as in the
abstract grouping. Again, the failure probability in the field ranks
higher. Hence, this factor cannot be fixed without changing the output
variance significantly. A further observation is that some of the
setup costs can be set to fixed value what reduces the measurement effort.

\subsection{Discussion and Consequences}

From the observations above we can conclude that the labour costs, the
sequence of technique application and the removal costs of most
techniques
are not an important part of the model and the variation in effort
does not have strong effects on the output, i.e., the ROI in our case.
On the other hand, the type of artifact or document the defect is
contained in, the difficulty of defect detection, and the removal
costs in the field have the strongest influences.

This has several implications: (1) We need more empirical research
on the distribution of defects over different document types and
the removal costs of defects in the field to
improve the model and confirm the importance of the factor, (2) we
still need more empirical studies on the effectiveness of different
techniques as this factor can largely reduce the output variance,
(3) the labour costs do not have to be determined in detail and it
does not seem to be relevant to reduce those costs, (4) further 
studies on the sequence and the removal costs are not necessary.

\section{Practical Application}
\label{sec:practical}

As we discussed above, the theoretical model can be used for analyses
but is too detailed for a practical application. The main goal is, 
however, to optimise the usage of \DDT s which requires a
applicability in practice. Hence, we need to simplify the model to
reduce the needed quantities.

\subsection{General}

For the simplification of the model, we use the following 
additional assumptions:

\begin{itemize}
  \item Faults can be categorised in useful defect types.
  \item Defect types have specific distributions regarding
        their detection difficulty, removal costs, and failure probability.
  \item The linear functional form of the difficulty approximates
        all other functional forms sufficiently.
\end{itemize}

We define $\TYPE$ to be the defect type of fault $i$. It is determined
using the defect type distribution of older projects.
In this way we do not have to look at individual
faults but analyse and measure defect types for which the determination
of the quantities is significantly easier. 
In the practical model we assumed that the defects can be grouped in
``useful'' classes or defect types. For reformulating the equation
it was sufficient to consider the affiliation of a defect to a type
but for using the model in practice we need to further elaborate on
the nature of defect types and how to measure them.

For our economics model
we consider the defect classification approaches from IBM \cite{kan02}
and HP \cite{grady92} as most suitable because they are proven to be
usable in real projects and have a categorisation that is coarse-grained
enough to make sensible statements about each category.

We also lose the concept of defect propagation as it was shown not to
have a high priority in the analyses above but it introduces significant
complexity to the model. Hence, the practical model can be simplified
notably.

\subsection{Equations}

Similar to Sec.~\ref{sec:equations_ideal} where we defined the basic
equations of the ideal model, we formulate the equations for the
practical model using the assumptions from above.

\subsubsection{Single Economics}

We start with the direct costs of a \DDT . Now we do
not consider the ideal quantities but use average values for the cost
factors. We denote this with a bar over the cost name.

\begin{equation}
  \label{eq:direct_practical}
  \CDIRECT = \MSETUP + \MEXEC
             + \sum_{i}{
             (1 - \DIFFT) \MREMV},
\end{equation}
where $\MSETUP$ is the average setup cost for technique $A$, $\MEXEC$ is
the average execution cost for $A$ with length $\LENGTH$, and $\MREMV$
is the average removal cost in defect type $\TYPE$. Apart from using
average values, the main difference is that we consider defect types
in the difficulty functions. The same applies to the
revenues.

\begin{equation}
  \label{eq:saved_practical}
  \REV =
  \sum_i{\pi_{\tau_i} (1 - \DIFFT)(\MREMVF + \MEFF)},
\end{equation}
where $\MEFF$ is the average effect costs of a fault of type $\TYPE$.
Finally, the future costs can be formulated accordingly.
\begin{equation}
  \label{eq:future_practical}
  \CFUT =
  \sum_i{\pi_{\tau_i} \DIFFT (\MREMVF + \MEFF)}.
\end{equation}

With the additional assumptions, we can also formulate a unique
form of the difficulty functions:
\begin{equation}
\theta_A(\tau_i, t_a) = m t_A + 1,
\end{equation}
where $m$ is the (negative) slope of the straight line. If a
technique is not able to detect a certain type, we will set
$m = 0$.

\subsubsection{Combined Economics}

Similarly, the extension to more than one technique can be done.

\begin{equation}
\begin{split}
\label{eq:practical_direct_combined}
  d_X = \sum_{x \in X}{\biggl[ \MSETUPX + \MEXECX} + 
       \sum_i{
       (1 - \theta_x(\TYPE,t_x))}\\
       \prod_{y < x}{\Bigl( \theta_y(\TYPE,t_y)}
       \Bigr) \MREMVX  \biggr]
\end{split}
\end{equation}

\begin{equation}
  t_X = \sum_i{\pi_{\TYPE} \prod_{x \in X}{\Bigl (\theta_x(\TYPE,t_x)
                  \Bigr)
                  \Bigl( \MREMVF + \MEFF \Bigr) }}
\end{equation}

\begin{equation}
\begin{split}
\label{eq:practical_revenues_combined}
  r_X = \sum_{x \in X}{\sum_i{\pi_{\TYPE}
       (1 - \theta_x(\TYPE,t_x))}} \\
        \prod_{y < x}{\Bigl( \theta_y(\TYPE,t_y)}
                             \Bigr)
        \Bigl( \MREMVF + \MEFF \Bigr) 
\end{split}
\end{equation}

\subsection{Sensitivity Analysis}

Similar to the analyses in Sec.~\ref{sec:sensitivity} we determined
the first and total order indexes of the practical model again with
data from \cite{wagner:tumi06,Wagner2006}. The results
are shown in Tab.~\ref{tab:practical} with the first order indexes left
and the total order indexes right. We have to note that we only looked
at defects in the code because we have no empirical data on defect
types in other kinds of documents. Furthermore, we introduced the
factor $\alpha$ that denotes the fraction of defects of a specific
defect type. 

\begin{table}[htbp]
\caption{The first and total order indexes from the practical model}
\begin{center}
\begin{tabular}{|l|r|l|r|}
\hline
$t$ & 0.1196 & $t$ & 0.8855 \\ \hline
$\pi$ & 0.1138 & $v_f$ & 0.8670 \\ \hline
$\bar{\theta}$ & 0.1097 & $s$ & 0.7881 \\ \hline
$\alpha$ & 0.0975 & $\bar{\theta}$ & 0.7857 \\ \hline
$v_f$ & 0.0694 & $l$ & 0.7772 \\ \hline
$l$ & 0.0634 & $\alpha$ & 0.6676 \\ \hline
$s$ & 0.0592 & $\pi$ & 0.6200 \\ \hline
$u$ & 0.0476 & $u$ & 0.4902 \\ \hline
$v$ & 0.0018 & $v$ & 0.0958 \\ \hline
\end{tabular}
\end{center}
\label{tab:practical}
\end{table}

We see that the effort for the techniques ranks highest in both
settings. The failure probability again ranks high in the factors
priorisation setting. Hence, this factor should be investigated in
more detail. Similarly to the ideal model, the setup and removal costs
of the techniques do not contribute strongly to the output variance.

In the factors fixing setting, we see that the setup and removal costs
can be fixed without changing the variance significantly. This implies
that we can use coarse-grained values here. Also the failure probability
can be taken from literature values. More emphasis, however, should be
put on the effort, the removal costs in the field, and the sequence of
technique application. Of which the last one is surprising as for the ideal
model this factor ranked rather low.

\subsection{Optimisation}
\label{sec:optimisation}

For the optimisation only two of the three components of the model
are important because the future costs and the revenues are dependent
on each other. There is a specific number of faults that have associated
costs when they occur in the field. These costs are divided in the two
parts that are associated with the revenues and the future costs, respectively.
The total always stays the same, only the size of the parts varies
depending on the used defect-detection techniques. Therefore, we use only the 
direct costs and the revenues
for optimisation and consider the future costs to be dependent on the
revenues.

Therefore, the optimisation problem can be stated by: maximise $r_X - d_X$.
By using Eq.~\ref{eq:practical_direct_combined} and 
Eq.~\ref{eq:practical_revenues_combined} we get the following equation to
be maximised.
\begin{equation}
\begin{split}
\sum_x{\biggl[ - \MSETUPX - \MEXECX + \sum_i{
(1 - \theta_x(\TYPE,t_x)) }} \\
\prod_{y < x}{(\theta_y(\TYPEJ,t_y))} 
\bigl(\pi_{\TYPE} \MREMVF + 
\pi_{\TYPE} \MEFF -
 \MREMVX \bigr)
\biggr]
\end{split}
\end{equation}

The equation shows in a very concise way the important factors in the
economics of \DDT s. For each technique there is the fixed setup cost and
the execution costs that depend on the effort. Then for each fault in the
software (and over all fault classes) we use the probability that the technique 
is able to
find the fault and no other technique has found the fault before to 
calculate the expected values of the other costs. The revenues are the
removal costs and effect costs in the field with respect to the failure
probability because they only are relevant if the fault leads to a failure.
Finally we have to subtract the removal costs for the fault with that
technique which is typically much smaller than in the field.

For the optimisation purposes, 
we probably also have some restrictions, for example a maximum effort
$t_{\textit{max}}$ with $\sum_x{t_x} \leq t_{\textit{max}}$, either fixed
length or none $t_A = \{0,100\}$, or some
fixed orderings of techniques, that have
to be taken into account. The latter is typically true for different forms
of testing as system tests are always later in the development than unit
tests.

Having defined the optimisation problem and the specific restrictions
we can use standard algorithms for solving it. It is a hard problem
because it involves multi-dimensional optimisation over a permutation
space, i.e., not only the length of technique usage can be varied but also
the sequence of the \DDT s.

\subsection{Applications}
\label{sec:applications}

In this section, we describe two possibilities how the practical model
can be used.

We can use the model in experiments as well as during normal software
development. As discussed in Sec.~\ref{sec:discussion} we can use the
ideal model to explain the effects of techniques on the economics. The
practical model is suited to measure important aspects of \DDT s in
software engineering experiments by finding difficulty functions of
certain techniques in certain domains. In software projects, the practical
model can also help to optimise the future quality assurance by using
the information from old projects.

\subsubsection{In-House}

The main idea is to predict the future
economics based on the data from finished projects.
The approach should then contain the following parts:
\begin{itemize}
  \item Classify found faults
  \item Which technique found which fault?
  \item Which faults were found in the field?
  \item Estimate failure probability and costs for each fault
\end{itemize}
From this data, we can estimate the needed quantities. 
This estimation process can have different forms.
The failure probability can either be estimated by expert opinion or using
field data if it was a field failure. The cost data can be partly taken
from effort measurements during development and from the field. Then
we can try to answer the two questions:
What is the optimal length of a technique and what is the optimal combination?
However, not that the results are in all cases dependent on the problem 
class and domain because they have a huge influence on the costs.

\subsubsection{Domain-Specific}

A second application could be to try to generalise the results of the
model to a complete domain either from field studies or experiments. 
There are probably specific defect types in
specific domains for which we might be able to collect data that is not
only valid inside one company but for the whole domain. In this way,
data from other companies could be used for optimisation purposes.

\section{Related Work}
\label{sec:related}

Our own previous work on the quality economics of \DDT s forms the
basis of this model. We formulated some simple relationships of cost
factors and how this could be used in evaluating and comparing different
techniques in \cite{wagner:sew05}. This is refined in \cite{wagner:wosq05}
and additional means to predict future costs are incorporated.
Some first results of the current model and
sensitivity analysis can be found in \cite{wagner:wosq06}.

The available related work can generally classified in two categories:
(1) theoretical models of the effectiveness and efficiency of either
test techniques or inspections and (2) economic-oriented, abstract
models for quality assurance in general. The first type of models is
able to incorporate interesting technical details but are typically
restricted to a specific type of techniques and often economical
considerations are not taken into account. The second type of models
typically comes from more management-oriented researchers that
consider economic constraints and are able to analyse different types
of defect-detection but often contain the technical details in a
very abstract way.

Pham describes in \cite{pham00} various flavours of a software cost model
for the purpose of deciding when to stop testing. It is a representative
of models that are based on reliability models. The main problem with such
models is that they are only able to analyse system testing and no other \DDT s
and the differences of different test techniques cannot be considered.

Holzmann describes in \cite{holzmann01} his understanding of the economics
of software verification. He describes some trends and hypotheses that
are similar to ours and we can support most ideas although they
need empirical justification at first.

Kusumoto et al.\ describe in \cite{kusumoto92,kusumoto93} a metric for
cost effectiveness mainly aimed at software reviews. They introduce the
concept of virtual software test costs that denote the testing cost that
have been needed if no reviews were done. This implies that we always want
a certain level of quality.

A model similar to the Kusomoto model but with an additional concept
of defect propagation was proposed by Freimut et al.\ in \cite{freimut05}.

The economics of the inspection process are investigated in \cite{biffl01}.
This work also uses defect classes and severity classes to determine the
specific costs. However, it identifies only the smaller removal costs to
be the benefit of an inspection.

An example of theoretical models of software testing is the
work of Morasca and Serra-Capizzano \cite{morasca04}. They concentrate
on the technical details such as the different failure rates. In
this paper there is also a detailed review of similar models.

Ntafos describes some considerations on the cost of software
failure in \cite{ntafos98}. The difficulties of collecting appropriate
data are shown but the model itself is described only on an
abstract level.

In \cite{krishnan96,slaughter98} a metric called
\emph{return on software
quality (RO\-SQ)} is defined. It is intended to financially justify investments in
quality improvement. The underpinnings of this metric are similar to
the analytical model defined in this paper although there are significant
differences. Firstly, it aims mainly on measuring the effects of process
improvements, i.\,e.\ constructive quality assurance, whereas we concentrate
on analytical quality assurance. Secondly, they base the calculations
mainly on average defect content in the software and do not consider
the important question if the faults lead to failures.

In \cite{knox93} the model of software quality costs is set into relation
to the Capability Maturity Model (CMM) \cite{paulk95}. The emphasis is hence on
the prevention costs and how the improvement in terms of CMM levels
helps in preventing failures.

Galin extends in \cite{galin04a,galin04b} the software quality costs
with managerial aspects
but the extensions are not relevant in the context of \DDT s.

Guidelines for applying a quality cost model in a business environment
in general are given in \cite{Kaner1996}.
Mandeville describes in \cite{Mandeville1990} also software quality costs,
a general methodology for cost collection, and how specific data from
these costs can be used in communication with management.

Humphrey presents in \cite{humphrey95} his understanding of software
quality economics. The defined cost metrics do not represent monetary
values but only fractions of the total development time. Furthermore,
the effort for testing is classified as failure cost instead of
appraisal cost.

Collofello and Woodfield propose in \cite{collofello89}
a metric for the cost efficiency but do not incorporate failure
probabilities or difficulties.

Based on the general model for software cost estimation COCOMO, the
COQUALMO model was specifically developed for quality costs in \cite{chulani99}.
This model is different in that it is aiming at estimating the costs
beforehand and that it uses only coarse-grained categories of \DDT s. In
our work, we want to analyse the differences between techniques in more
detail.

Boehm et al.\ also present in \cite{boehm04} the iDAVE model that uses
COCOMO II and COQUALMO. This model allows a thorough analysis of the ROI
of dependability. The main difference is again the granularity. Only an
average cost saving per defect is considered. We believe that analysing
costs per defect type can improve estimates and predictions.

Building on iDAVE, Huang and Boehm propose a value-based approach for
determining how much quality assurance is enough in \cite{huang05}. In
some respect that work is also more coarse-grained than our work because
it considers only the defect levels from COQUALMO. However, it contains
an interesting component that deals with time to market costs that are
currently missing from our model.

A somehow similar model to COQUALMO in terms of the description of
the defect introduction and removal process is described in \cite{jalote03}.
However, it offers means to optimise the resource allocation. The only
measure for \DDT s used is defect removal efficiency.

\section{Conclusions}
\label{sec:conclusions}

We finally summarise our work and the main contributions and give
some directions for future work.

\subsection{Summary}

We propose an analytical model of quality economics with a strong
focus on \DDT s. This focus is necessary to be able to be more detailed
than comparable approaches. In this way, we incorporate different cost
types that are essential for evaluating \DDT s and also a notion of
reliability or the probability of failure. The latter is also very important
because it is significant which faults are found in terms of reliability.
This distinguishes the
model from more abstract approaches. On the other hand we have models
derived from software reliability modelling. These models are typically
simpler but can only be used on techniques where reliability models can
be applied, i.\,e. mainly system tests. We aim to incorporate all types
of \DDT s.

One of the main contributions is also the research priorisation. We find
that it is most rewarding to further investigate the distribution of
defect over document types, the removal costs in the field, and the
difficulty, especially the functional form with respect to varying effort.
All of these have not been subject to extensive empirical work.

The main weakness of our model is that the ideal model is not usable
in real software projects. Hence, we derived a practical model that
is based on defect types. This gives us a greater data basis for each
type. The problem here is, that it is not totally clear if this structuring
in defect types really is able to give useful distributions of the
removal costs, removal difficulty, and failure probability. Furthermore, it
strongly depends on how ``good'' these types are defined and we currently
have no requirements on the classes.

\subsection{Future Work}

As future work, we consider working on support for the estimation of
the needed quantities of the practical model, especially the number 
of faults $\bar{I}$ and also on the probability of failure of the
defects as those are important factors. An application of the model
to a real project and thereby analysing the predictive validity of
the model is one of next major steps.

The optimisation must be worked on in more
detail and effective tool support is essential to make the model
applicable in practice. Finally, an incorporation of time to market might be
beneficial because there are important costs associated with time overruns
that need to be considered. In some markets this may be even more important
than all the other factors contained in the model.

%ACKNOWLEDGMENTS are optional
\section{Acknowledgments}
We are grateful to Sandro Morasca for detailed comments on the model and
to Bev Littlewood for helping on the understanding of their diversity
model. This research was supported by the \emph{Deutsche
Forschungsgemeinschaft (DFG)} within the project
\emph{InTime}.

\bibliographystyle{abbrv}

% ACM needs 'a single self-contained file'!

% \appendix

% \section{Difficulty Values}

\balancecolumns

\end{document}